# Tracking the circulation routes of fresh coins in Bitcoin: A way of identifying coin miners with transaction network structural properties

Zeng-Xian Lin[1], Xiao Fan Liu[2*]

(1 School of Computer Science and Engineering, Southeast University, Nanjing 211189
2 Department of Media and Communication, City University of Hong Kong, Hong Kong SAR)
* xf.liu@cityu.edu.hk

**Abstract:** Bitcoin draws the highest degree of attention among cryptocurrencies, while coin mining is one of the most important fashion of profiting in the Bitcoin ecosystem. This paper constructs fresh coin circulation networks by tracking the fresh coin transfer routes with transaction referencing in Bitcoin blockchain. This paper proposes a heuristic algorithm to identifying coin miners by comparing coin circulation networks from different mining pools and thereby inferring the common profit distribution schemes of Bitcoin mining pools. Furthermore, this paper characterizes the increasing trend of Bitcoin miner numbers during recent years.

**Keyword:** Bitcoin; blockchain; transaction network; pooled mining; coin miners;

## 0 Introduction

In 2008, Satoshi Nakamoto published the article "Bitcoin: A Peer-to-Peer Electronic Cash System," which proposed a decentralized virtual electronic currency scheme based on peer-to-peer transactions [1]. The Bitcoin system was then launched in January 2009. As of 2017, its market value exceeded 100 billion US dollars, and the transaction records exceeded 3 billion US dollars. The success of Bitcoin's technical and economic experiments has attracted widespread attention from scholars. Public Bitcoin transaction records were analyzed in order to describe the Bitcoin ecosystem. In 2014, Kondor found that the growth rate of hub nodes in the Bitcoin trading network was faster than that of nodes with lower degrees, and summarized the preferential attachment phenomenon of "the rich become richer" in the Bitcoin trading network [2]. In view of the anonymous nature of Bitcoin, Reid and Fleder proposed several heuristic methods to de-anonymize Bitcoin addresses in 2013 and 2015, respectively [3][4]. In 2016, Sapirshtein extended the underlying model of selfish mining attacks, and provided an algorithm to find attackers in the model [5]. In 2017, Easley studied the role of Bitcoin transaction fees in the evolution of the blockchain from a mining-based structure to a market-based ecology [6].

Mining was the most direct way to gain benefits in the Bitcoin ecosystem. In 2015, Lewenberg et al. proposed a theoretical model for the formation and reward allocation mechanism of Bitcoin mining pool teams [7]. In 2017, Beccuti clarified the impact of Bitcoin mining mechanism on miners' honesty [8]. Adoption of the Proof of Work mechanism in the Bitcoin system guaranteed the security and fairness of mining and rewarded the system maintainers (i.e., miners) who paid computing power by issuing tokens [1]. The computing power paid by miners was proportional to the probability of gaining rewards. However, when the number of maintainers in the Bitcoin ecosystem was increased significantly, the rewards became difficult to be obtained on the basis of the computing power of a single miner. Therefore, individual miners joined the mining pool one after another, and received stable remuneration from the mining pool according to the proportion of their contributed computing power. After the first mining pool Eligius was formed in 2011, as of the end of 2012, the total computing power of all mining pools in the world reached 50% of the entire network. Nowadays, the total computing power of all mining pools was even stabilized at more than 95% of the entire network [9][10]. Currently, four of the top five mining pools in computing power are located in China, whose computing power has exceeded 51% of the entire network [10]. Many are concerned that once the security of the system is attacked by several jointly large mining pools, the Bitcoin system may be seriously threatened.

In view of this concern, the number of miners in the entire network must be clearly determined, and the distribution of each miner's contribution to computing power is extremely valuable for further understanding the economic ecology of Bitcoin. This paper proposed a method to identify miner groups under various mining pools from Bitcoin transaction records, and the revenue distribution model of Bitcoin mining pools was summarized. First, the trend of new coins in each block after generation was tracked, and a new coin circulation network was built. Second, this study compared the new coin circulation network constructed in the blocks generated by the same mining pool and different mining pools at adjacent times. By analyzing the proportion, location, and reasons of the intersection and difference of the network, the miner groups in the mining pool were located in the network, and the mining benefit distribution mode of the mining pool was summarized. Finally, the changes in the size of the miner group over the years were also analyzed.

# 1. Miners, mining pools, and revenue distribution

"Mining" refers to the process in which system maintainers in the Bitcoin ecosystem update the Bitcoin ledger. The Bitcoin ledger can be immutable by adding a random number to the updated data block of the ledger and by enabling the hash value of the data block to fall into a specific range. The process in which Bitcoin system maintainers find the correct random number is mining. Pool mining is a multi-user agreement that stipulates that multiple clients jointly completed the hash calculation on the basis of the same data block. The mining rewards are finally distributed to the clients according to the proportion of the contributed computing power by the mining pool. Clients who joined the mining pool are collectively referred to as miners.

Miners are generally provided with common income distribution methods by Bitcoin mining pools, such as Pay per Last N Shares (PPLNS), Pay per Share (PPS), and SOLO. In the PPLNS model, "revenue is paid according to the past N shares." If the mining pool can successfully mine, then the miners will be given dividends according to the sum of the computing power contributions provided by the miners in the past period of time after several blocks are confirmed. If the mining pool did not dig a new block, then the miner will not receive any dividends. In the PPS model, revenue is paid according to the miners' computing power. This set-up prevents the situation wherein the income is very high and no income will be given. According to the proportion of the miners' computing power in the mining pool, regardless of whether the mining pool digs new blocks or not, dividends equal to the computing power contributed by the miners will be distributed. The mining pools generally issue rewards at a fixed time every day. In the SOLO mode, a miner mines alone. If a new block was mined, then the mining pool will distribute all the rewards to the single miner [8].

Identification of miners in Bitcoin transaction records is difficult due to the different revenue distribution patterns of different mining pools. In this paper, from the perspective of a complex network, the circulation of new coins in a block was tracked, and the Bitcoin capital flow network was established. The miner group of the mining pool was found by comparing the distribution paths of rewards obtained by different mining pools.

## 2. Establishment of the new coin circulation network

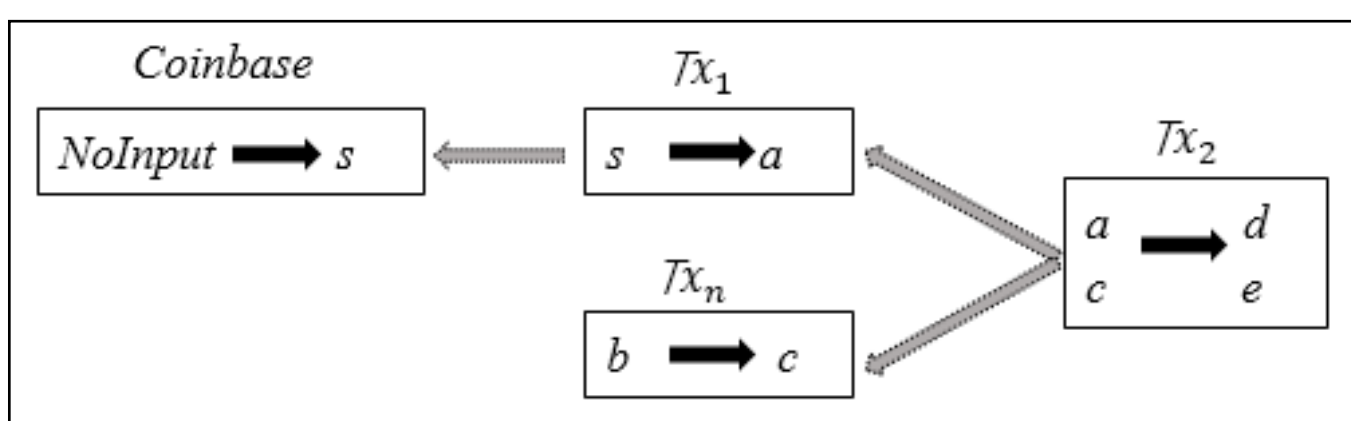

Fig.1 referencing of Bitcoin transaction

Figure 1 shows a schematic diagram of a series of typical transaction records in the Bitcoin ledger. Each rectangular box represents a transaction, the solid arrow is the direction of the fund transfer in the transaction, and the dotted arrow represents the reference relationship of the transaction. In each transaction, bitcoin funds are transferred from the input payment address on the left to the output receiving address on the right. The bitcoins spent by the input address in each transaction shall be referenced from the output address in the previous transaction, and an output address can only be referenced once. For example, input addresses *a* and *c* in transaction $Tx_2$ were referenced from the output addresses *a* and Bitcoin *c* in transactions $Tx_1$ and $Tx_n$, respectively. *Coinbase* transaction was the first transaction in each block and was used to record the miner's mining reward. It only had the billing address of the output, but not the input address.

In this paper, the input address of *Coinbase* was recorded as *NoInput*, and output address *s* was the receiving address (mining address) for mining.

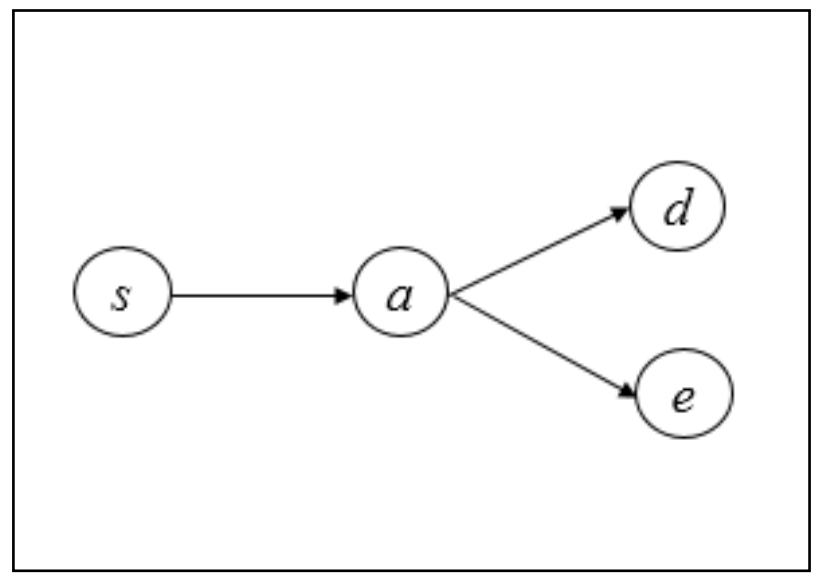

Fig.2 schematic diagram of fresh coin circulation network

The new coin circulation network established in this paper referred to a network composed of nodes whose output addresses were directly or indirectly referenced by the *Coinbase* transactions of each block within a certain period of time. A directed edge can be found between the output address of the previous transaction and each output address of the next transaction. Figure 2 shows a new coin circulation network established according to the transaction relationship in Figure 1. Bitcoin addresses *b* and *c* cannot be found in the new coin circulation network as they did not directly or indirectly refer to mining address *s*.

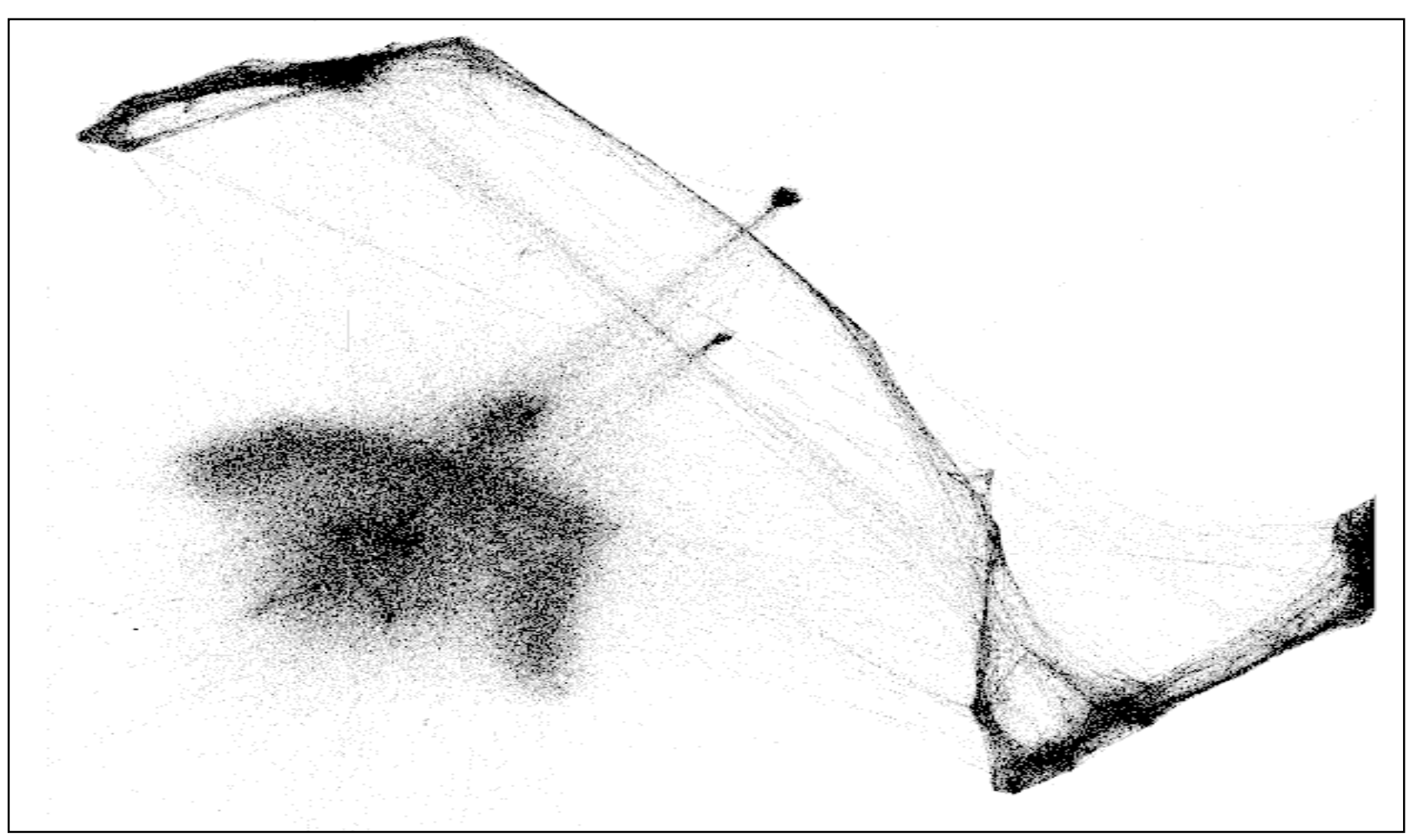

Fig.3 a fresh coin circulation network

As revenue was generally distributed to miners by the mining pools within a period of time when new coins were mined, the traceable transactions generated by the new bitcoin within seven days were selected to build a complete new coin circulation network in this paper. Figure 3 shows a typical new coin circulation network. This new coin was mined in the 277937th block of the Bitcoin transaction record. A total of 96,714 transactions were tracked in four days. Moreover, a total of 168,999 nodes and 693,037 edges can be found in the constructed network. The density of the network was almost 0. Calculated through the undirected graph, the average degree of nodes was only 4.1, and the clustering coefficient was 0.024. As shown in Figure 4, the degree distribution of the network was in a multimodal long-tail form, indicating the existence of hub nodes in the network.

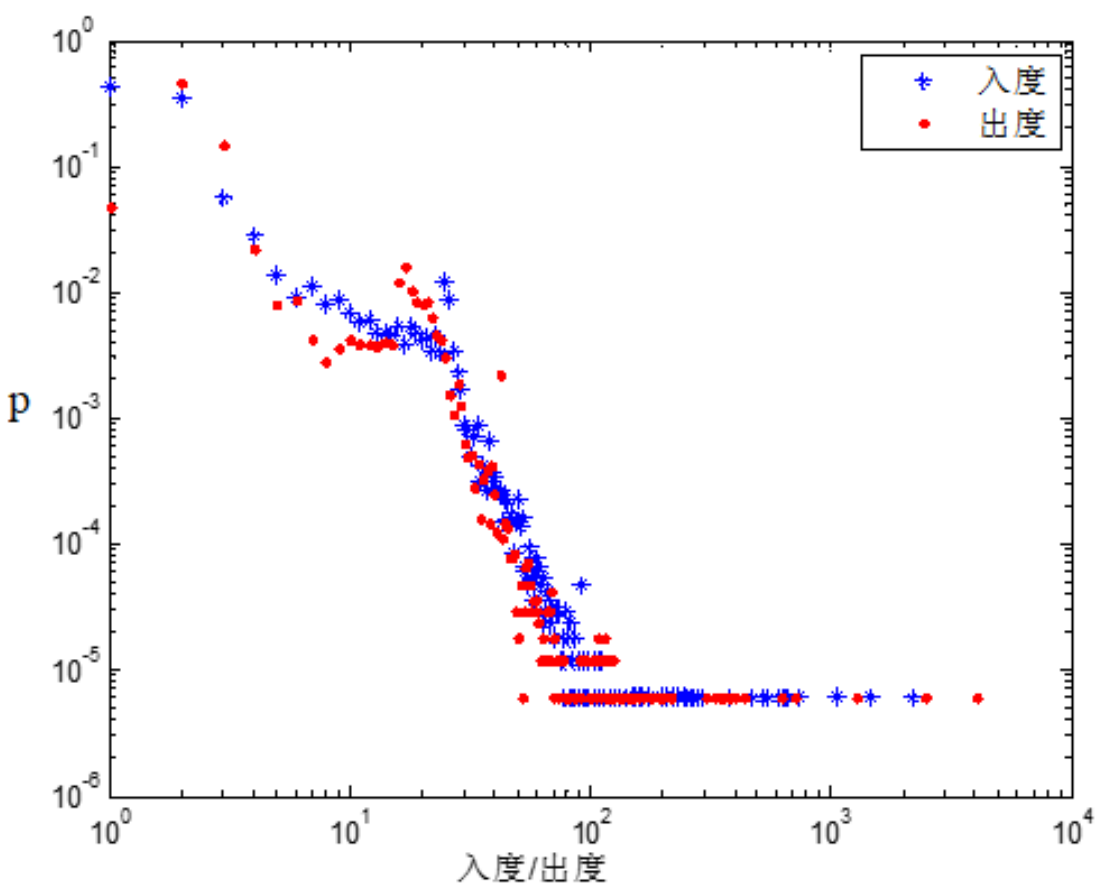

Fig.4 in- and out- degree distributions of the network

## 3. Miner group identification

After new coins were mined, the mining pool will transfer the new coins to the miners' personal addresses according to the proportion of their contribution through a certain income distribution model. Generally, miners will only contribute all their computing power to a certain mining pool and only obtain profit distribution from a single mining pool. Moreover, the collection address submitted by the miners to the mining pool will not be changed in a short period of time. Therefore, the miner group can be realized by comparing the circulation network of new coins mined by different mining pools, that is, when two new coins are selected with adjacent time distances in the Bitcoin transaction record, the new coins may flow to the same group of miners if the two new coins were mined by the same mining pool. If the two blocks are from different mining pools, then the new coins may flow to different miner groups.

In this paper, the new coin circulation network $G_1 = \{V_1, E_1\}$ and $G_2 = \{V_2, E_2\}$ were first established on the basis of mining addresses $s_1$ and $s_2$ in blocks 277940 and 277941, respectively, on the blockchain dug by the BTC Guild mining pool. As shown in Table 1, through a comparison of $G_1$ and $G_2$, we can find that the two networks were similar in scale and structure, and the network node sets $V_1$ and $V_2$ were almost the same. Afterwards, the new coin circulation network $s_3$ and $s_4$ were established on the basis of mining addresses $G_3 = \{V_3, E_3\}$ and $G_4 = \{V_4, E_4\}$ in block 277937 mined by the GHash.IO mining pool and block 277938 mined by the CloudHashing mining pool. The comparison shows that the size of the two networks were different, in which $V_4$ contained most of the nodes in $V_3$ and also contained more than 30% of the nodes that did not appear in $V_3$.

Tab.1 Comparison of two consecutive fresh coin circulation networks

| New coin circulation | $G_1$ | $G_2$ | $G_3$ | $G_4$ |
|---|---|---|---|---|
| Starting block number | 277,940 | 277,941 | 277,937 | 277,938 |
| Mining pool | BTC Guild | GHash.IO | CloudHashing | |
| Transaction count | 343,925 | 344,240 | 344,396 | 344,238 |
| Traceable transactions | 180,387 | 179,386 | 96,714 | 147,451 |
| Number of network | 302,799 | 301,191 | 168,998 | 250,451 |
| Number of overlapping | 299,359 | 168439 | | |
| Proportion of | 98.9% | 99.4% | 99.7% | 67.2% |
| Number of network | 1,471,647 | 1,477,792 | 765,327 | 1,160,988 |
| Longest network path | 456 | 456 | 270 | 456 |

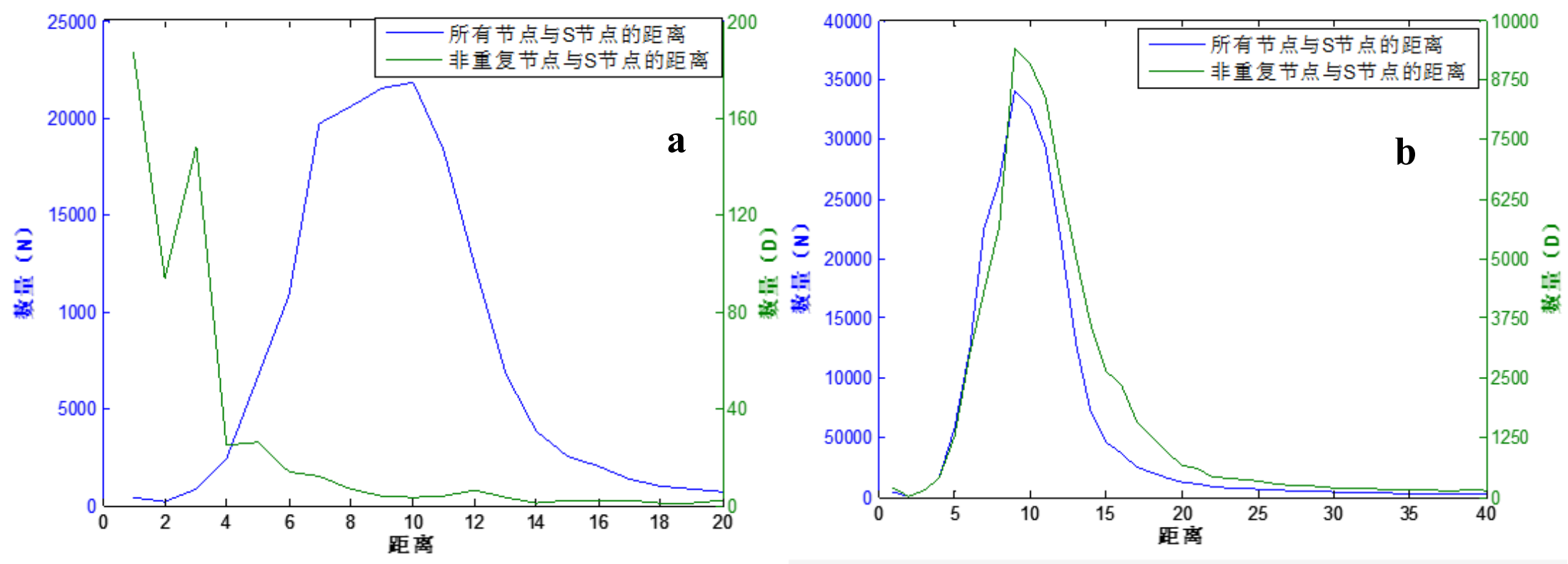

Fig.5 the distance distributions from all nodes to source nodes *s* of the network (a) $G_3$ network (b) $G_4$ network

The network structure properties of the two pairs of non-overlapping nodes in the new coin circulation network $G_3$ and $G_4$ were further analyzed, and the distances between network source nodes $s_3$ and $s_4$ in both networks were examined. As shown in Figure 5, most of the distances between all nodes in the network and the source node were distributed around 10 hops, and the distance distribution of non-repeated nodes was slightly different from that of all nodes.

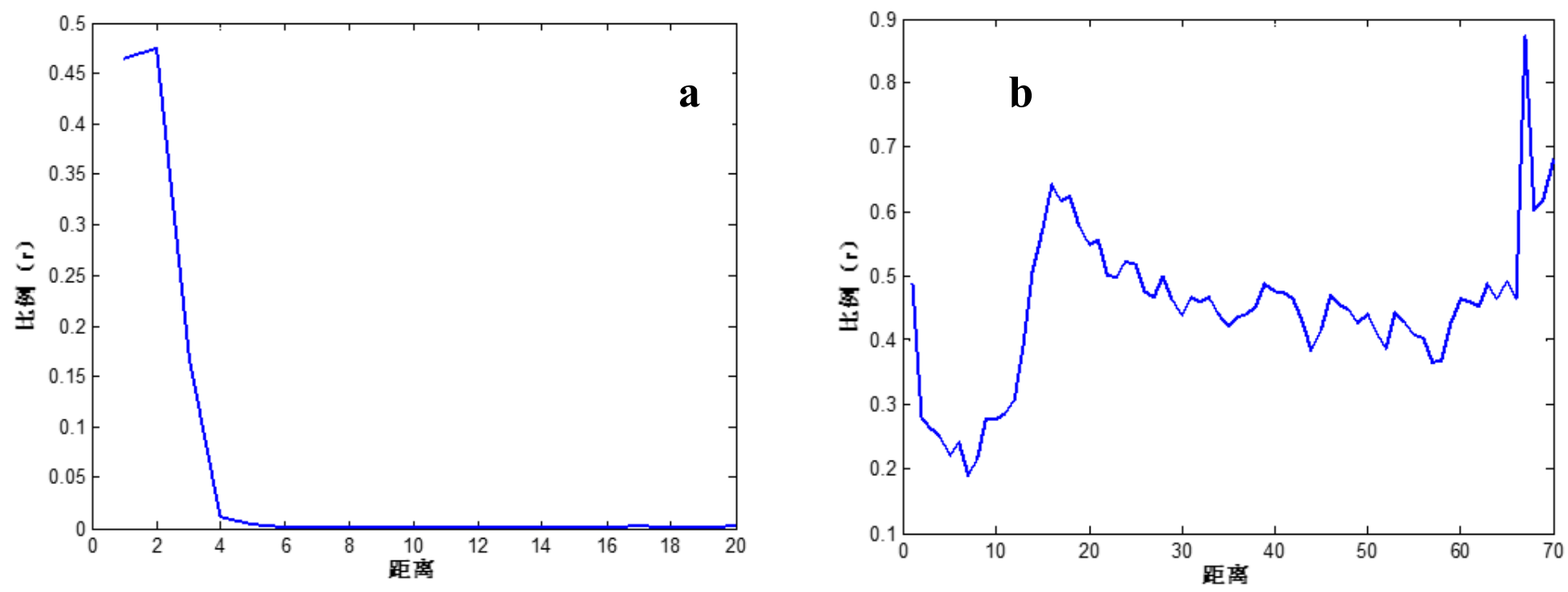

Fig.6 proportions of non-overlapping nodes by their distance to the source nodes (a) $G_3$ network (b) $G_4$ network

The number of node sets $\boldsymbol{n}_i$ with a length of *i* from source node *s* in network *G* was denoted by $N_i$, and the number of non-repeated node sets $\boldsymbol{d}_i$ was denoted by $D_i$. Under different path lengths, the proportion of non-repeated nodes to the number of all nodes is:

$$r_i = \frac{D_i}{N_i}. \tag{2}$$

Figure 6 shows the proportion of non-repeated nodes to all nodes in the $G_3$ and $G_4$ networks. If the two selected new coins were from the same mining pool, then the nodes close to the network source node *s* were completely duplicated, and they can be determined to be the same group of miners. If the two mining pools were selected from different blocks, then the probability of non-repeated nodes appearing near the source node was higher, and these non-repeated nodes can be considered miners of the mining pool. Figure 7 shows the schematic diagram of identifying miner groups through adjacent new coin circulation networks, where *s* was the mining address and *a*, *b*, *c*, and *d* were miner addresses.

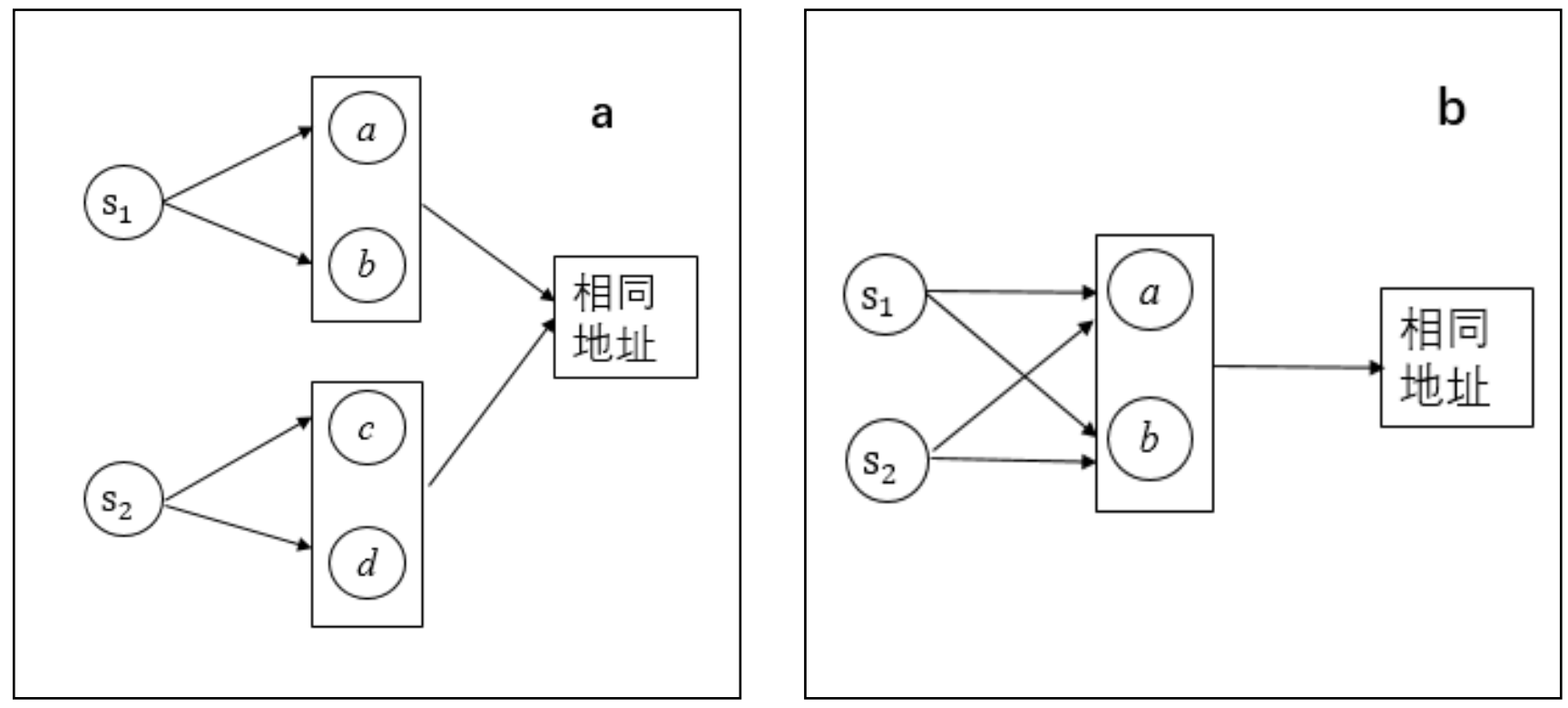

Fig.7 consecutive fresh coin circulation network schematic diagram. (a) different mining pools, (b) same mining pool.

## 4. Inference of mining pool revenue distribution model

According to Section 1, different revenue distribution methods can be found in different mining pools after mining new blocks. The miner reward patterns of major mining pools were analyzed and summarized into three types through the miner identification methods described in Sections 2 and 3.

**Mode 1: Indirect distribution**

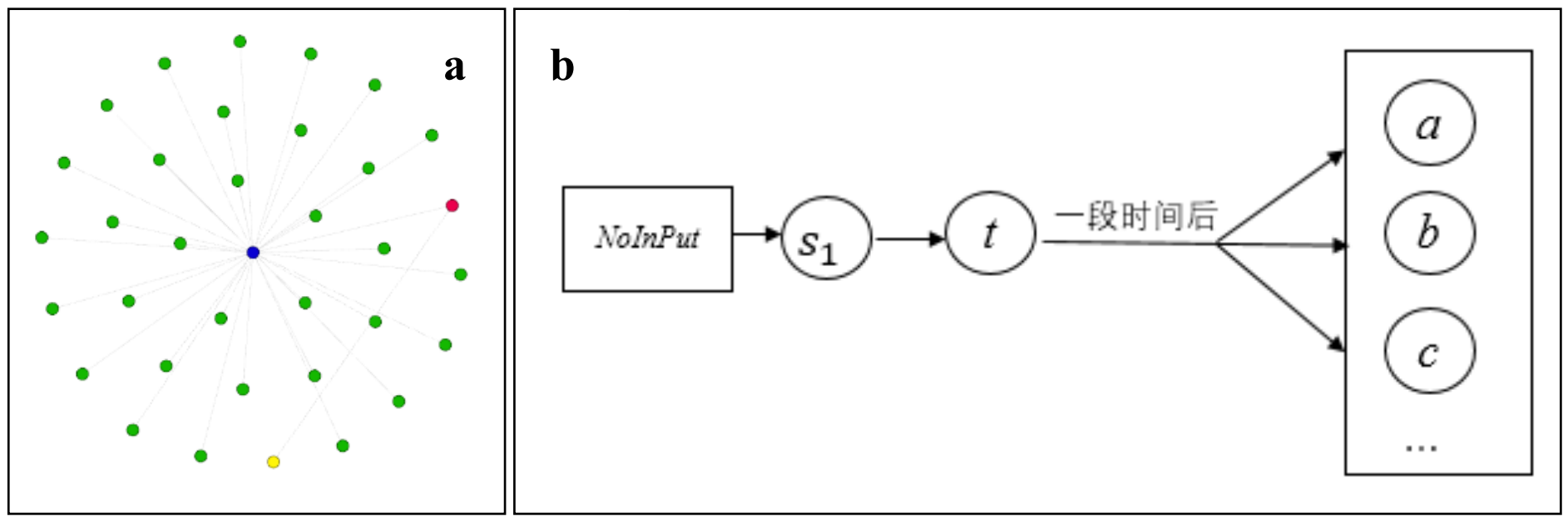

Fig.8 Pattern 1: indirect distribution. (a) coin circulation network, (b) schematic diagram

As shown in Figure 8(a), the yellow node shows the *NoInput* address, the red node is the mining address *s* of the mining pool, the blue node is the intermediate address used by the mining pool, and the green node is the group of miners. In Mode 1, when the mining pool mined a new block, the mining revenue was first allocated to its own mining address by constructing a *Coinbase* transaction in the new block. Then, the mining revenue was transferred to an intermediate address, and after a period of time, the mining revenue will be uniformly distributed to the miners by the intermediate address. The representative mining pool of this mode was Ozcoin.

**Mode 2: Direct distribution**

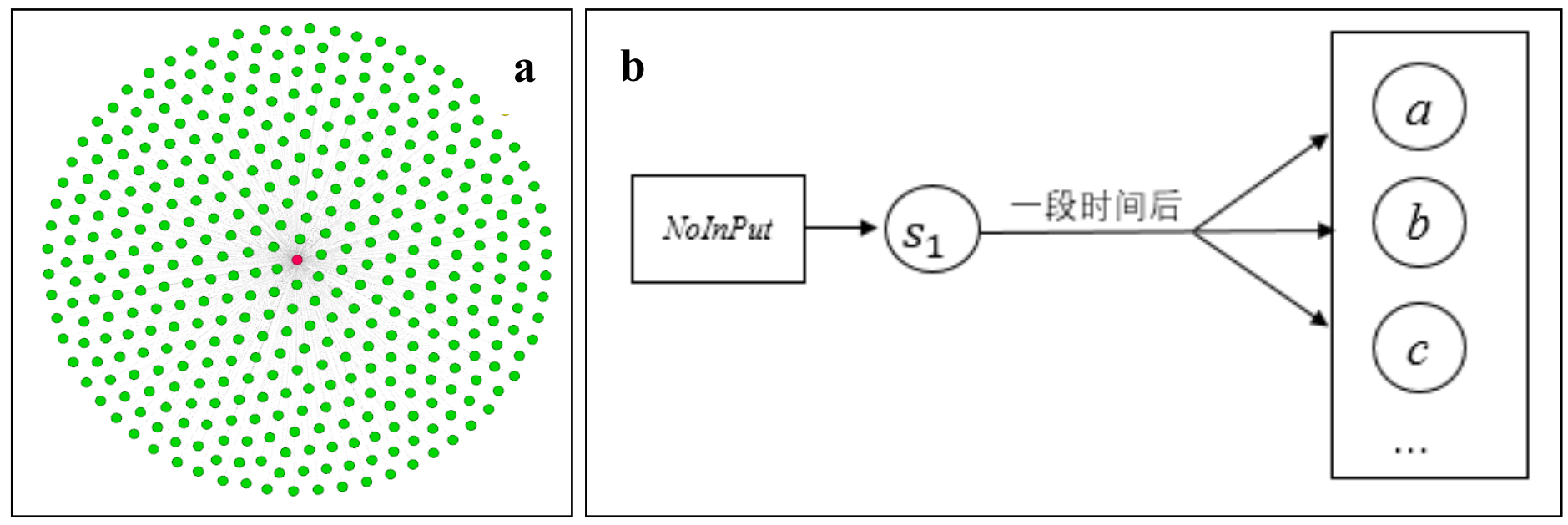

Fig.9 Pattern 2: direct distribution. (a) coin circulation network, (b) schematic diagram

As shown in Figure 9(a), the yellow node shows the *NoInput* address, the red node is the mining address *s* of the mining pool, and the green node is the group of miners. In Mode 2, when the mining pool mined a new block, the mining revenue was first allocated to its own mining address by constructing a *Coinbase* transaction in the new block. After a period of time, the mining revenue will be uniformly distributed to the miners. The representative mining pool of this mode was GHash.IO.

**Mode 3: Immediate distribution**

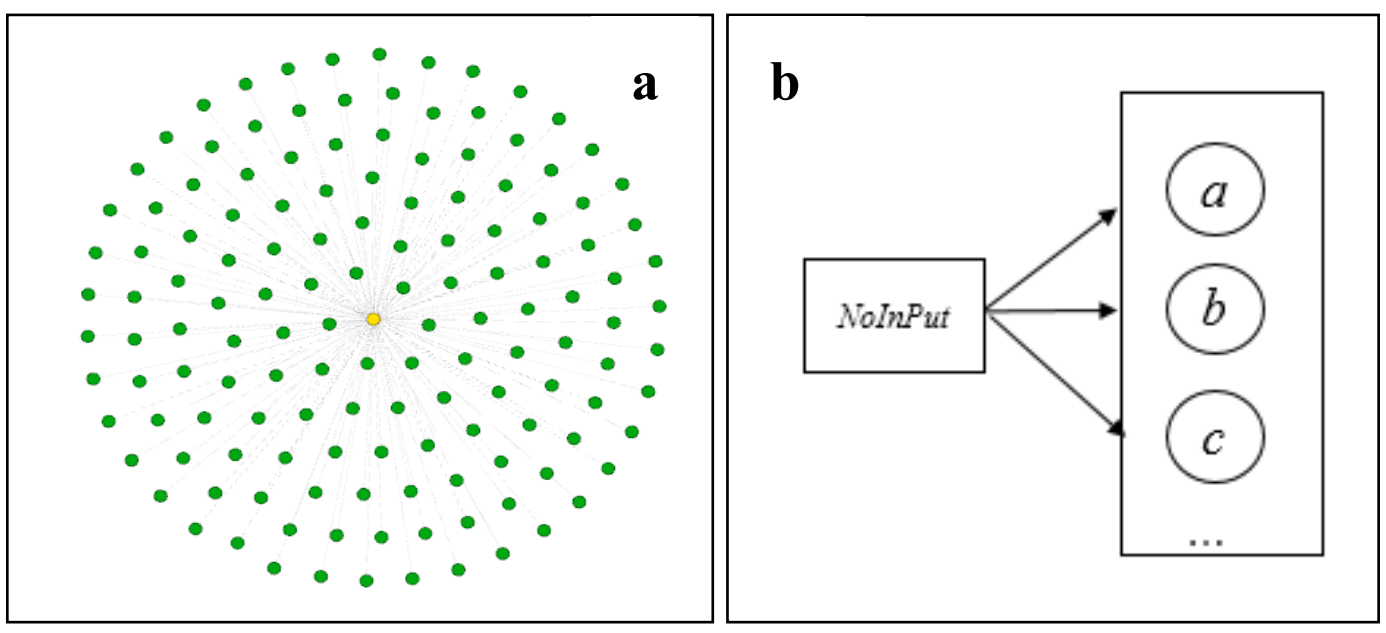

Fig.10 Pattern 3: immediate distribution. (a) coin circulation network, (b) schematic diagram

As shown in Figure 10(a), the yellow node shows the *NoInput* address, and the green node is the group of miners. In Mode 2, when the mining pool mined a new block, the mining revenue was distributed to the miners directly in the *Coinbase* transaction in the new block. This mode was more direct than the first two modes. The representative mining pool of this mode was Eligius.

## 4. Analysis of the number of miners over the years

The Bitcoin mining pool Eligius was formed in 2011. By the end of 2011, the total computing power of all mining pools was less than 20% of the total network computing power. To date, the total computing power of all mining pools in the world has reached more than 95% of the total computing power of the entire network. In this section, the miner groups were searched in the mining pools with the most mining from 2012 to 2016 through the methods proposed in Sections 2 and 3.

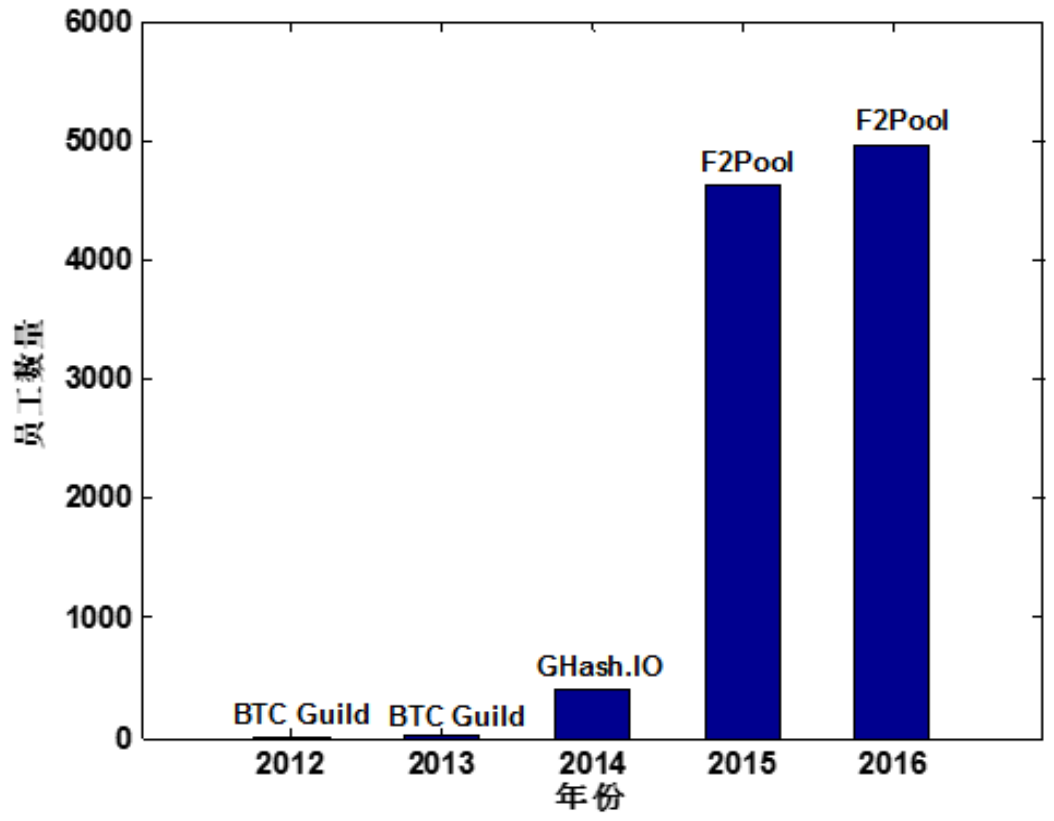

Fig.11 the scale of miners' group in the most active mining pool from 2012 to2016

Figure 11 shows the changes in the number of miners in the world's largest mining pools from 2012 to 2016. The number of miners was increasing year by year from more than a dozen miners in 2012 to nearly 5,000 in 2016. In the current Bitcoin ecosystem, profiting from collective mining has become an ecological consensus. This change was far from Satoshi Nakamoto's original intention of "one CPU, one vote" for the Bitcoin system [1].

# 5. Discussion

This study proposed a method to identify Bitcoin miners. The circulation network of new coins was established by tracking the movement of new coins in Bitcoin blocks over the course of a week. A comparison of the differences between the time-adjacent new coin circulation networks of different mining pools showed that the new coin circulation networks of different mining pools had a large number of non-repeated nodes at the end close to the network source. Moreover, the node repetition rate was gradually decreased and tended to be stable with the distance between the nodes and the network source. Thus, non-repeated nodes close to the origin of the network were considered the miner addresses. Regular or irregular revenue distribution can be performed for the group of miners identified by this method and the mining pools they belong to. Three basic mining pool revenue distribution models were also summarized on the basis of Bitcoin transaction records. Finally, this study described the year-by-year growth trend of miner group size. The number of Bitcoin miners greatly increased in 2015, which meant that the mining ecology of the Bitcoin system gradually moved away from its original design.

Identifying Bitcoin miners was of great help in further inferring the computing power of each miner. In today's ecological background where mining computing power is gradually concentrated, it was only necessary to control a limited number of mining pools or even a limited number of miners with huge computing power so that it can pose a substantial threat to the security of the system, and the Bitcoin system was increasingly at risk of being attacked by 51%. The further work in this study will be focused on the evaluation of the computing power of a single miner in order to make a more accurate judgment on the Bitcoin system.

# References

[1] Satoshi Nakamoto. Bitcoin: A peer-to-peer electronic cash system[J]. Consulted, 2008.

[2] Kondor D, Pósfai M, Csabai I, et al. Do the rich get richer? An empirical analysis of the Bitcoin transaction network.[J]. PloS one, 2014, 9(2):e86197.

[3] Reid F, Harrigan M. An Analysis of Anonymity in the Bitcoin System[C]. Barcelona, Spain:

IMC’13,2013:1318 - 1326.
[4] Fleder M, Kester M S, Pillai S. Bitcoin Transaction Graph Analysis[J]. arXiv:1502.01657, 2015.
[5]Sapirshtein A, Sompolinsky Y, Zohar A. Optimal Selfish Mining Strategies in Bitcoin[C]. Springer, Berlin, Heidelberg: International Conference on Financial Cryptography and Data Security, 2016:515-532.
[6] Easley D, O'Hara M, Basu S. From Mining to Markets: The Evolution of Bitcoin Transaction Fees[J]. Available at SSRN: https://ssrn.com/abstract=3055380, 2017.
[7] Lewenberg Y, Bachrach Y, Sompolinsky Y, et al. Bitcoin Mining Pools: A Cooperative Game Theoretic Analysis[C]. Istanbul, Turkey: International Conference on Autonomous Agents and Multiagent Systems, 2015:919-927.
[8] Beccuti J, Jaag C. The Bitcoin Mining Game: On the Optimality of Honesty in Proof-of-work Consensus Mechanism[J]. Working Papers, 2017.

[9] http://www.8btc.com/bitcoin-mining-pool

[10] https://btc.com/stats/pool